\documentclass[aps,prb,reprint,superscriptaddress,floatfix,longbibliography]{revtex4-2}

\usepackage{amsmath}
\usepackage{amssymb}
\usepackage{bm}
\usepackage{braket}
\usepackage{graphicx}
\usepackage{booktabs}
\usepackage{multirow}
\usepackage{xcolor}
\usepackage[
  colorlinks=true,
  citecolor=blue,
  linkcolor=blue,
  urlcolor=blue
]{hyperref}

\newcommand{\be}{\begin{equation}}
\newcommand{\ee}{\end{equation}}
\newcommand{\bea}{\begin{eqnarray}}
\newcommand{\eea}{\end{eqnarray}}

\begin{document}
\title{P-wave Magnetism and Gate-Tunable Edelstein Response \\in van der Waals Heterostructures}

\author{Hanbyul Kim}
\address{Department of Physics, Hanyang University, Seoul 04763, Republic of Korea}

\author{Chan Bin Bark}
\address{Department of Physics, Hanyang University, Seoul 04763, Republic of Korea}

\author{Seik Pak}
\address{Department of Physics, Hanyang University, Seoul 04763, Republic of Korea}

\author{GiBaik Sim}
\email{gibaik.sim@unimelb.edu.au}
\address{Department of Physics, University of Melbourne, Grattan Street, Parkville VIC 3010, Australia}

\author{Moon Jip Park}
\email{moonjippark@hanyang.ac.kr}
\address{Department of Physics, Hanyang University, Seoul 04763, Republic of Korea}
\address{Research Institute for Natural Science and High Pressure, Hanyang University, Seoul, 04763, Republic of Korea}
\date{\today}

\begin{abstract}
Odd-parity magnetism has attracted significant interest for its unconventional spin splitting. However, a concrete microscopic route for its realization remains elusive. In this work, we propose van der Waals heterostructures of stripe antiferromagnets (sAFMs) as an ideal platform for electrically controllable $p$-wave magnetism. In the sAFM/metal/sAFM structure, the leading RKKY-type exchange interaction is canceled due to the symmetry of the stacking pattern. This exposes a higher-order biquadratic interaction as a dominant contribution that drives a filling-controlled transition from a collinear phase to an orthogonal $p$-wave configuration. The resulting $p$-wave phase exhibits a gate-tunable Edelstein response, which originates from magnetic symmetry breaking rather than conventional relativistic spin-momentum locking and remains robust even under substantial spin-orbit coupling. Finally, we propose material candidates for the realization of our theory. Our results establish van der Waals heterostructures as a practical platform for non-relativistic spintronics with electric control of $p$-wave spin textures.
\end{abstract}

\maketitle

\section{Introduction}  Momentum-dependent spin splitting is conventionally associated with relativistic spin-orbit coupling (SOC)~\cite{Rashba, RashbaLikephysics}. However,
recent progress in the field of unconventional magnetism has demonstrated that significant spin splitting can arise without relativistic effects~\cite{Naka2019,Smejkal2020,Hayami2019JPSJ,Yuan2020PRB,Smejkal2022}. A representative class of such materials is altermagnets~\cite{Mazin2022, Krempasky2024, Feng2022,Smejkal2022AMR, Gonzalez2021PRL, Betancourt2023,liu2025altermagnetism, Han2024SciAdv}. Despite vanishing net magnetization, the intertwining of magnetic order with crystalline symmetries yields a large exchange-driven spin splitting with even-parity spin texture. The strong spin splitting driven by exchange couplings has positioned such unconventional magnets as a promising platform for a wide range of spintronics applications~\cite{Bose2022,Bai2022,Karube2022,Smejkal2022Rev, Shao2021,Mohsen1,Mohsen2,Mohsen3,zhao2026multiferroic,wang2026aus}.

On the other hand, odd-parity spin textures have been proposed in noncollinear magnets~\cite{Hayami2020PRB, Yuan2021PRM, Watanabe2024, ntsh-ypmc, hu2025spin}. The odd-parity phase can occur when the magnetic order is intertwined with crystalline translations through nonsymmorphic symmetry. A prototypical example is a $p$-wave magnet, whose dispersion exhibits $E_\uparrow(\mathbf{k})=E_\downarrow(-\mathbf{k})$, indicating that the spin polarization at $\mathbf{k}$ is opposite to that at $-\mathbf{k}$~\cite{Song2025,yamada2025metallic, hellenes2023p,zk69-k6b2,chakraborty2025,85fd-dmy8,7cpd-mv33,ydjj-6r9l,c8pd-2fs4,li2026dynamical,li2026p,habel2026magnetic,han2026optical}. A spin-helix state has been proposed as a candidate realization~\cite{Song2025,yamada2025metallic}, and bulk materials such as CeNiAsO have been proposed to host the requisite symmetry settings~\cite{hellenes2023p,zk69-k6b2,chakraborty2025}.

 Nevertheless, recent angle-resolved photoemission spectroscopy (ARPES) measurements on CeNiAsO reported no resolvable $p$-wave spin splitting~\cite{CeNiAsO_ARPES1, CeNiAsO_ARPES2}, leaving the existence of p-wave magnetism in this system a subject of ongoing debate. While general symmetry criteria for odd-parity itinerant antiferromagnetism have been established based on space-group and Wyckoff-position analysis~\cite{zk69-k6b2,85fd-dmy8}, a concrete material realization that satisfies these criteria and simultaneously exhibits sizable exchange-induced spin splitting has not been identified. It is therefore desirable to identify material platforms that realize p-wave magnetism in a controllable fashion.

Here, we propose a magnetic trilayer heterostructure composed of two stripe antiferromagnets (sAFMs) separated by a metallic spacer (sAFM/metal/sAFM) as a promising platform for $p$-wave magnetism. The metallic spacer mediates a filling-tunable interlayer exchange that favors an orthogonal configuration of the two stripe order parameters, thereby generating a $p$-wave spin texture controllable by electrostatic gating. Importantly, the metallic layer enables electrical driving and transport readout of the $p$-wave spin splitting configuration via the Edelstein response with conduction electrons directly exchange-coupled to the stripe order parameters.
Recent advances in van der Waals (vdW) magnets~\cite{Deng2018, Burch2018, Gibertini2019, Chu2020, Ni2021, Kim2019, Kim2021} make this proposal experimentally realistic. In particular, vdW $\mathrm{GdTe}_3$ retains stripe magnetism down to the monolayer limit~\cite{GdTe3AFM,GdTe3monolayer,GdTe3VdW}. Finally, we propose a gate-tunable Edelstein response as a smoking-gun electrical signature of the $p$-wave phase.

\begin{figure*}[t]
\centering
\includegraphics[width=0.9\linewidth]{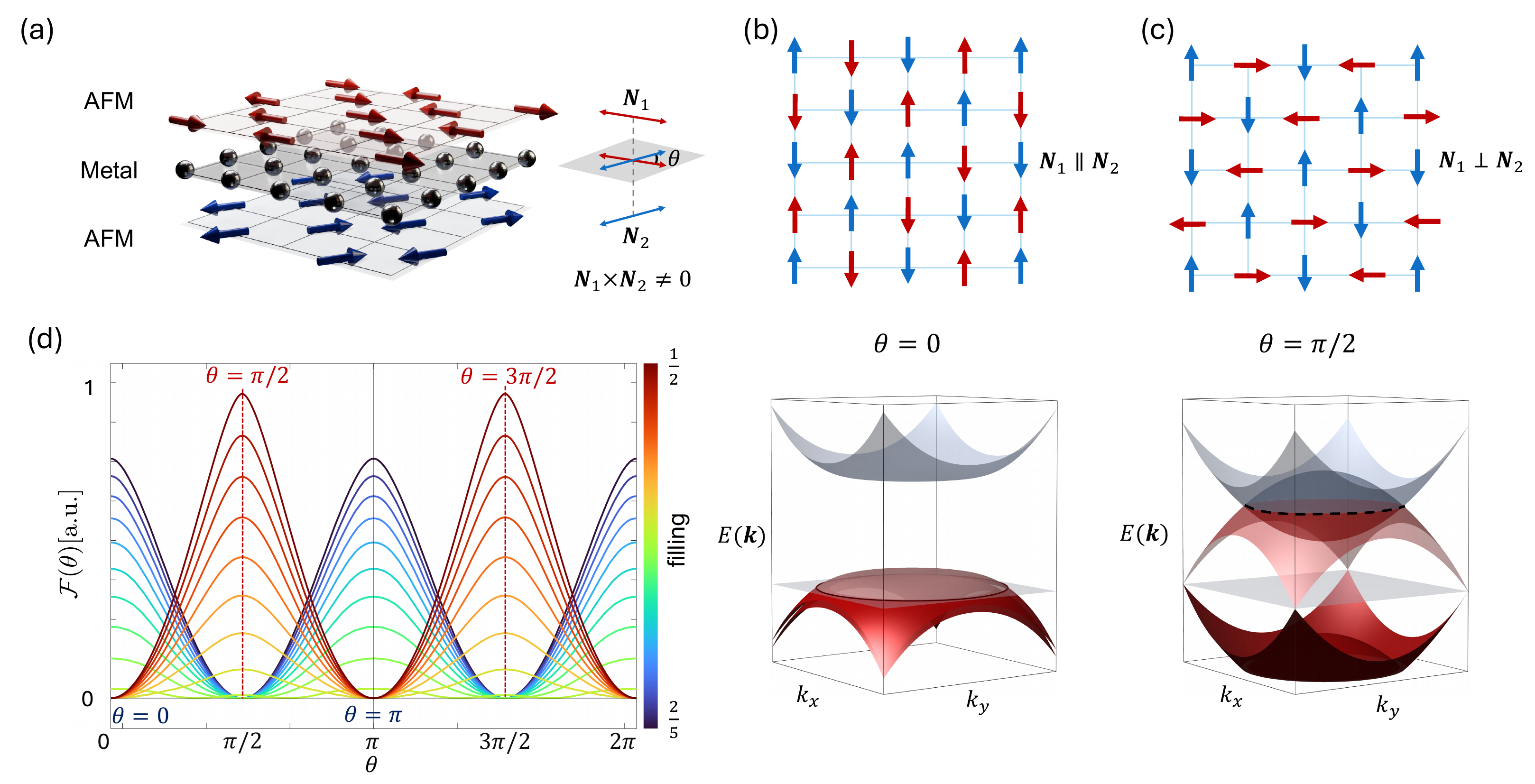}
\caption{(a) Schematic of the sAFM/metal/sAFM heterostructure.  $\mathbf{N}_{1}$ and $\mathbf{N}_{2}$ form a relative in-plane angle $\theta$.
(b) Band structure for the collinear state ($\theta=0$), where $\mathcal{PT}$-symmetry enforces global spin degeneracy.
(c) Band structure for the orthogonal state ($\theta=\pi/2$). The breaking of $\mathcal{PT}$-symmetry lifts the degeneracy, resulting in $p$-wave band splitting. (d) Relative free energy $\mathcal{F}(\theta)$ as a function of $\theta$ for different electronic fillings of the metallic layer. The color gradient indicates filling levels from $n=1/3$ (blue) to $n=2/5$ (red). The system undergoes a transition from an orthogonal $p$-wave phase ($\theta=\pm\pi/2$) at low filling to a collinear phase ($\theta=0, \pi$) at high filling.} 
\label{fig:schemaric}
\end{figure*}

\section{Phenomenological model} We consider the two independent stripe order parameters $\mathbf{N}_1$ and $\mathbf{N}_2$ for the top and the bottom layers respectively [See Fig.~\ref{fig:schemaric}(a)], where the local moment in each layer ($\lambda=1,2$) is parameterized as
$
\mathbf{m}_\lambda(\bm r)= \mathbf{N}_\lambda \cos\!\big[\mathbf{Q}\cdot(\bm r-\bm r_\lambda)\big],
$
where $\mathbf{Q}$ is the ordering wavevector of the stripe state, $\bm r_\lambda$ is the layer-dependent registry vector. In the absence of SOC, symmetry considerations allow us to write the free-energy density in a general form as~\cite{Ivanov2009, PhysRevB.98.245129, zk69-k6b2},

\begin{align}
\mathcal{F} &=
\alpha(T)\sum_{i=1,2}|\mathbf N_i|^2
+\beta_1\!\left(\sum_{i=1,2}|\mathbf N_i|^2\right)^{\!2}
+v|\mathbf N_1|^2|\mathbf N_2|^2  \nonumber\\
&\quad +\beta_2(\mathbf N_1\cdot\mathbf N_2)^2
+J_{\rm eff}(\mathbf N_1\cdot\mathbf N_2)
+\cdots .
\end{align}

Here, $\alpha(T)$ governs the intrinsic magnetic stability. The coefficients $\beta_1$ and $v$ describe the amplitude quartic terms of the two stripe order parameters, whereas $\beta_2$ controls the relative-angle-dependent biquadratic interaction. In particular, the term $v|\mathbf N_1|^2|\mathbf N_2|^2$ is symmetry-allowed and is generally independent of the coefficient $\beta_2$ of $(\mathbf N_1\cdot\mathbf N_2)^2$.
For the angular energetics discussed below, we fix the magnitudes of the order parameters,
$|\mathbf N_1|=|\mathbf N_2|\simeq N$. In this limit, the \(\beta_1\) and \(v\) terms give only an angle-independent constant, \(4\beta_1 N^4+vN^4\), and therefore do not affect the selection between collinear and orthogonal configurations.
The quartic free energy is bounded from below in the parameter regime of interest, $\beta_1>0$, $4\beta_1+v>0$, and $4\beta_1+v+\beta_2>0$, ensuring stability of both the collinear ($\theta=0$) and orthogonal ($\theta=\pi/2$) configurations since our parameter regime these conditions are satisfied (see Sec.~II of the Supplemental Material~\cite{SM} for details).
Consequently, up to an angle-independent constant, the effective free energy depends only on the relative angle $\theta$. The equilibrium configuration is governed by the competition between the linear and biquadratic exchange couplings,

\begin{align}
\mathcal{F}(\theta) \approx J_{\mathrm{eff}} N^2 \cos\theta + \beta_2 N^4 \cos^2\theta.
\end{align}
In our model, $J_\textrm{eff}$ vanishes due to the geometric cancellation. Microscopically, the effective exchange $J_\textrm{eff}$ is generated by the metallic layer via the Ruderman–Kittel–Kasuya–Yosida (RKKY)-type contribution to the free energy~\cite{Mohsen1}, $
\Delta \mathcal{F}^{(2)}
\sim -\sum_{\bm q} 
\mathbf{m}_{1}(- \bm q)\,\chi(\bm q)\mathbf{m}_{2}(\bm q)
\label{eq:DF2_tensor}
$, where $ \chi(\bm q)$ is the static spin-susceptibility tensor of the metal. For the stripe order, only the Fourier components at \(\bm q=\pm\mathbf{Q}\) contribute, giving
$
J_{\rm eff}\propto \chi(\mathbf{Q})\cos(\mathbf{Q}\cdot\ \bm\delta)
$ with \(\bm\delta=\bm r_2-\bm r_1\) is the registry shift of the two AFM. In our trilayer geometry, the half-translation stacking enforces \(\mathbf{Q}\cdot \bm\delta=\pi/2\),  and the bilinear exchange vanishes, \(J_{\rm eff}=0\), independent of microscopic details of \(\chi(\bm q)\). As a result, the symmetry imposes the cancellation of the linear exchange coupling. (See Sec.~III of the Supplemental Material for details.)

\begin{figure*}[t]
\centering
\includegraphics[width=1\linewidth]{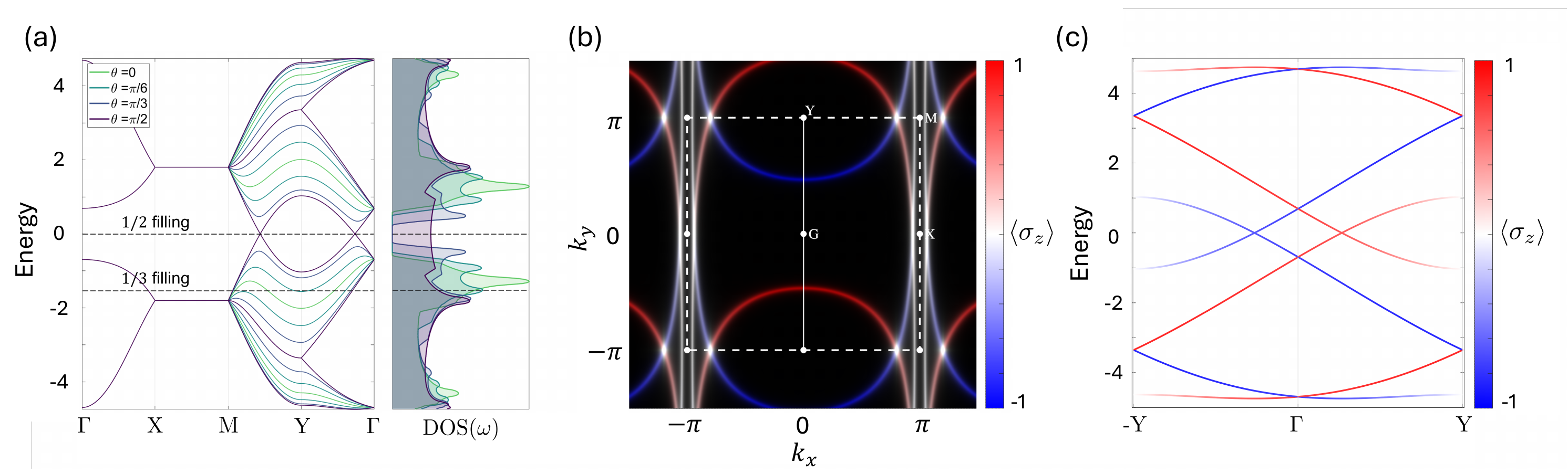}
\caption{
(a) Evolution of the Bloch bands and density of states (DOS) for different relative angles. The shaded DOS regions highlight the emergence of spectral splitting as $\theta$ approaches $\pi/2$.
(b) Momentum-space spin polarization map $\langle S_{z}(\mathbf{k})\rangle$ at the 1/3-filling. The texture exhibits the characteristic odd-parity symmetry, $\langle S_{z}(\mathbf{k})\rangle=-\langle S_{z}(-\mathbf{k})\rangle$.
(c) Spin-resolved band dispersion along the $Y\rightarrow -Y$ path [white solid line in (b)]. The crossing of opposite spin branches (red and blue) at $E_{F}$ (dashed line) demonstrates the symmetry-protected $p$-wave splitting without relativistic SOC. Computed from Eq. (3) with $t=1, m_1=m_2=1.8$; panel (a) at $\theta=0,\pi/6,\pi/3,\pi/2$ with a histogram DOS
}
\label{fig:band}
\end{figure*}

Single-ion magnetocrystalline anisotropy of the easy-plane form $F_{\rm aniso}^{(i)} = a(S_{i,x}^2 + S_{i,y}^2) + \beta S_{i,z}^2$, with $a<\beta$, as appropriate for the candidate sAFM materials considered here, confines the moments to the $xy$ plane and contributes only a $\theta$-independent constant to the in-plane angular free energy; it does not affect the collinear/orthogonal phase selection (see Sec.~V of the Supplemental Material). In contrast, the biquadratic term $\beta_2$ is insensitive to the inversion of the local moments, thus $\beta_2 (\mathbf{N}_1 \cdot \mathbf{N}_2)^2$ is non-vanishing leading-order contribution to the free energy. Depending on the sign of $\beta_2$, minimizing the free energy density $f \propto \beta_2 \cos^2\theta$ presents two distinct scenarios: a negative $\beta_2$ would stabilize the collinear alignment [$\theta = 0$, Fig.~\ref{fig:schemaric}(b)], whereas a positive $\beta_2$ favors the orthogonal alignment [$\theta = \pm \pi/2$, Fig.~\ref{fig:schemaric}(c)], corresponding to the $p$-wave magnetism with spin splitting. As we explicitly show in the next section, the change of electron filling alters the sign of $\beta_2$ [Fig.~\ref{fig:schemaric}(d)], where the minimum of $\mathcal{F}(\theta)$ shifts from the collinear configuration near half-filling to the orthogonal configuration at low filling .

Microscopically, the same trend is reflected in the electronic band structure with $\theta$, as schematically illustrated in Fig.~\ref{fig:schemaric}(b) and \ref{fig:schemaric}(c). For the collinear configuration ($\theta=0$), a gap opens at half-filling [Fig.~\ref{fig:schemaric}(b)]. 
This gap is induced by the staggered Zeeman field associated with the collinear phase analogous to a Slater-type insulating mechanism. 
The resulting gap formation lowers the electronic energy and therefore favors the collinear state, particularly at half-fillings.
 
Conversely, the orthogonal phase forms a symmetry-protected Dirac nodal line that forbids such global gap opening [Fig.~\ref{fig:schemaric}(c)]. In this metallic regime, the magnetic instability is driven by kinetic energy minimization. Crucially, the orthogonal configuration lifts the Kramers degeneracy inherent to the collinear phase, inducing a large spin splitting. Away from the half-filling, this splitting pushes the lower spin-branch band. By populating these energetically lowered states, the system can achieve a net reduction in total energy, stabilizing the $p$-wave state. 

\section{Lattice model of odd-parity phases} We implement the proposed magnetic transition by considering the tight-binding Hamiltonian of the metallic layers.
\begin{gather}
    \hat{H}_{\text{mid}} = -t \sum_{\langle i,j \rangle, \sigma} \left( c_{i, \sigma}^\dagger c_{j, \sigma} + \text{h.c.} \right) - \mu \sum_{i, \sigma} n_{i, \sigma} \nonumber \\
    + \sum_{i \in A} c_{i}^\dagger [\mathbf{m}_{\text{1}}(\mathbf{r}_i) \cdot \bm{\sigma}] c_{i}  
    + \sum_{j \in B} c_{j}^\dagger [\mathbf{m}_{\text{2}}(\mathbf{r}_j) \cdot \bm{\sigma}] c_{j},
    \label{eq:H_mid_compact}
\end{gather}
Here, $c_{i\sigma}^\dagger$ creates an electron with spin $\sigma$ at site $i$ in the middle layer, with $t$ and $\mu$ denoting the nearest-neighbor hopping and chemical potential, respectively. The last two terms describe the induced Zeeman potential. Here the top(bottom)-layer magnetization $\mathbf{m}_{\text{t(b)}}$ couples exclusively to the A-(B)-sublattice. The stripe spin texture is parametrized as
$\mathbf m_{\rm 1}(\mathbf r)=\mathbf N_1\cos(\mathbf Q\!\cdot\!\mathbf r)$ and
$\mathbf m_{\rm 2}(\mathbf r)=\mathbf N_2\cos(\mathbf Q\cdot\mathbf r)$ with $\mathbf Q=(\pi,0)$. Here, although we do not explicitly model the magnetic layers, Eq.~\eqref{eq:H_mid_compact} represents the effective Hamiltonian derived via a Schrieffer-Wolff (SW) transformation~\cite{BRAVYI20112793}, where integrating out the AFM layers generates the sublattice-selective exchange potentials. (See Sec.~II of the Supplemental Material for details of the SW calculation.)

Representative band structures derived from the effective model are shown in Fig.~\ref{fig:band}(a). At $\theta=0$, the electronic band retains the global spin degeneracy. This is due to the $\mathcal{PT}$-symmetry (combining time-reversal with spatial inversion) at the bond center. This symmetry enforces Kramers degeneracy throughout the Brillouin zone (BZ) and opens a global hybridization gap at half-filling. Any deviation from the collinear phase ($\theta\neq0$) breaks $\mathcal{PT}$-symmetry, lifting the spin degeneracy. Even for small relative angles, the electronic bands immediately acquire a characteristic $p$-wave-like odd-parity texture. As the angle approaches the orthogonal limit ($\theta\rightarrow \pi/2$), the spin splitting is maximized. 

On the other hand, exactly at $\theta=\pi/2$, we find the formation of the stable Dirac nodal line [Purple line in Fig.~\ref{fig:band}(b)]. This is due to the nonsymmorphic spin-rotation symmetry, $\mathcal{U}_\textrm{ns}=\{T_{1/2}^x\,|\,C^{S}_{4z}\}$, comprised of a half-translation($T_{1/2}^x$) and 90$^\circ$ spin-rotation($C^{S}_{4z}$) along $z$-axis. Each band is uniquely labeled by its eigenvalue under the non-symmorphic symmetry  $\mathcal{U}_\textrm{ns}$ throughout the BZ. The electronic states can be block-diagonalized into distinct sectors labeled by the eigenvalues. In particular, the bands crossing at the Fermi level belong to different sectors such that their hybridization is forbidden. The symmetry protection ensures that the orthogonal phase remains semimetallic at half-filling, in sharp contrast to the gapped collinear phase.

Upon doping (e.g., $n\sim1/3$), the energetic hierarchy between the two magnetic configurations reverses. Away from the gap, the instability is driven by the kinetic energy of itinerant carriers. While the collinear phase maintains spin-degenerate bands, the orthogonal phase ($\theta=\pi/2$) breaks $\mathcal{PT}$-symmetry, inducing a large momentum-dependent spin splitting. This splitting allows electrons to populate the energetically lowered spin-branch. The resulting kinetic energy gain overcomes the magnetic cost, stabilizing the orthogonal $p$-wave order in the metallic regime.  As shown in Fig.~\ref{fig:band}(b) at $n = 1/3$, the resulting metallic state exhibits an $p$-wave spin splitting texture in the out-of-plane direction, $\langle \sigma_z(\mathbf{k}) \rangle = -\langle \sigma_z(-\mathbf{k}) \rangle$ [Fig.~\ref{fig:band}(b),(c)]. The corresponding dispersion also shows that opposite-spin branches cross without avoided level crossing, demonstrating that the $p$-wave splitting is symmetry-protected. Further details of the Bloch Hamiltonian and the associated symmetry analysis are provided in Sec.~IV of the Supplemental Material.

\section{Perturbative analysis} The effective magnetic free energy can be derived by integrating out the itinerant electronic degrees of freedom. With the linear exchange geometrically suppressed, the leading magnetic interaction arises from fourth-order scattering processes, yielding the biquadratic coefficient $\beta_2$. Analytically, $\beta_2$ decomposes into two competing contributions, $\beta_2= \beta_2^{\rm FS} + \beta_2^{\rm sea}$,
\small
\begin{align}
\beta_2^{\rm FS} &=\! \sum_{\mathbf k} A_{\mathbf k}\big[\partial_\epsilon  n_\textrm{F}(\varepsilon_\mathbf{k}-\mu)+\partial_\epsilon  n_\textrm{F}(\varepsilon_\mathbf{k}+\mu)\big]\nonumber
\!+\!
(\mathbf k\rightarrow \mathbf k+\mathbf Q),\\
\beta_2^{\rm sea} &=\! -\sum_{\mathbf k} B_{\mathbf k}[ n_\textrm{F}(\varepsilon_\mathbf{k}-\mu)-n_\textrm{F}(-\varepsilon_\mathbf{k}-\mu)\big]
\!+\!
(\mathbf k\rightarrow \mathbf k+\mathbf Q),
\label{eq:beta}
\end{align}
\normalsize
where $n_\textrm{F}$ and $\partial_\epsilon n_\textrm{F}$ denotes the Fermi-Dirac distribution function and its derivative respectively. The explicit forms of the prefactors appearing in Eq.~(\ref{eq:beta}) are given by 
$A_\mathbf{k} = \frac{\lambda^4 \varepsilon_{\mathbf{k+Q}}^2}{16 \Delta_\mathbf{k}^2}$ and 
$B_\mathbf{k} = \frac{\lambda^4 \varepsilon_{\mathbf{k+Q}}^2}{16 \varepsilon_\mathbf{k} \Delta_\mathbf{k}^2} + \frac{ \lambda^4 \varepsilon_\mathbf{k} \varepsilon_{\mathbf{k+Q}}^2}{4\Delta_\mathbf{k}^3}$, 
with $\Delta_\mathbf{k} \equiv \varepsilon_\mathbf{k}^2 - \varepsilon_{\mathbf{k+Q}}^2$. (See Sec.~II of the Supplemental Material for the detailed derivation.) The Fermi-surface term restricts the contribution to the near Fermi surface, whereas the Fermi-sea term integrates over the entire BZ.  This decomposition highlights a competition between two distinct instability mechanisms.

\begin{figure}[t]
\centering
\includegraphics[width=1.0\linewidth]{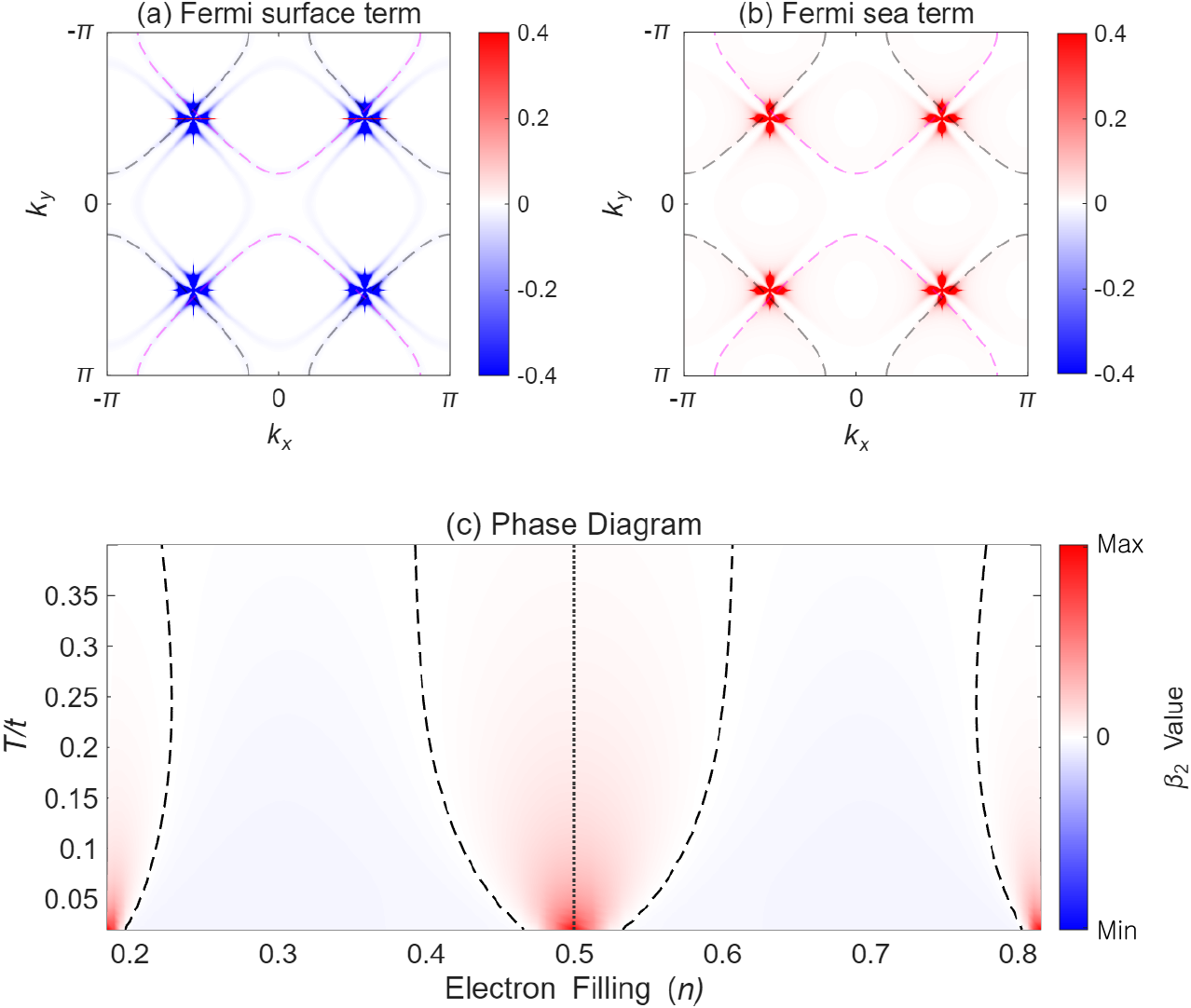}
\caption{
Microscopic origin of biquadratic exchange and magnetic phase diagram. (a–c) Momentum-resolved distribution of the $\beta_2$ integrand components : (a) Fermi surface contribution $(\propto n'_{F})$, dominating near the chemical potential; (b) Fermi sea contribution $(\propto n_{F})$, representing energy redistribution of all occupied states (c) Magnetic phase diagram in the filling $(n)$–temperature $(T/t)$ plane. The orthogonal $p$-wave phase $(\theta \approx \pi/2)$ is stabilized at low fillings by kinetic energy gain, whereas the collinear phase $(\theta = 0, \pi)$ is favored near half-filling $(n=1/2)$.
}
\label{fig:phase}
\end{figure}

From Eq. \eqref{eq:beta}, we find that the Fermi surface and sea term are positive and negative contributions, respectively. Interestingly, this feature is independent of microscopic band dispersions. As shown in Fig.~\ref{fig:phase}(a), the Fermi surface contribution is concentrated at hot spots in the BZ where the nesting condition $\varepsilon_\mathbf{k}=\varepsilon_{\mathbf{k+Q}}$ is satisfied. This term favors the orthogonal configuration by maximizing exchange-driven band splitting near the Fermi level analogous to the Stoner instability. In stark contrast, the second term represents the Fermi-sea contribution. As shown in Fig.~\ref{fig:phase}(b), this term contributes negatively ($\beta_2<0$) and is delocalized across the occupied states. It favors the collinear alignment to open a global hybridization gap, lowering the total energy of the Fermi sea.

Consequently, the magnetic transition is governed by a filling-controlled crossover between these two forces. Fig.~\ref{fig:phase}(c) maps out the resulting phase diagram. Near half-filling, the Fermi-sea contribution dominates due to the gap opening, stabilizing the collinear insulator. However, with doping, the Fermi-surface hot spots dominate the interaction, driving the transition to the orthogonal $p$-wave metal. This confirms that the $p$-wave magnetism is a tunable outcome of Fermi surface engineering.

\section{Edelstein effect}
We propose a gate-tunable Edelstein response as an electrical fingerprint of the $p$-wave phase.~\cite{manchon2019current, vzelezny2017spin, offidani2017optimal,Mohsen2} Within the constant relaxation-time approximation~\cite{li2015intraband,freimuth2014spin,gonzalez2024non}, we define the current-induced spin accumulation by a dimensionless Edelstein kernel $\chi_{ij}$ in the lattice as
\bea
\frac{\delta S_i}{\hbar/2}
\simeq 
\chi_{ij}\,
\frac{e E_j a \tau_{\rm sc}}{\hbar},
\label{eq:edelstein_scaling}
\eea
where $a$ is the in-plane lattice constant and $\tau_{\rm sc}$ is its momentum relaxation time. Here $\delta S_i$ denotes the spin polarization per magnetic unit cell, induced by an electric field $E_j$.

In our system, the out-of-plane spin polarization induced by a $y$-directed electric field is governed by
\begin{gather}
\chi_{zy}\!=\!
\sum_n \int_{\rm BZ}\!\frac{d^2k}{(2\pi)^2}\!
\langle\sigma_z\rangle_{n\bm{k}}\,
\langle\partial_{k_y}H(\bm{k})\rangle_{n\bm{k}}
\Bigl(\!-\partial_{\varepsilon} n_\textrm{F}(\varepsilon_{n\mathbf{k}}\!-\!\mu)\!\Bigr),
\nonumber
\\
\label{eq:chi_kernel_def}
\end{gather}
where $\langle O\rangle_{n\bm{k}}\equiv \langle n\bm{k}|O|n\bm{k}\rangle$, and $|n\bm{k}\rangle$ and $\varepsilon_{n\bm{k}}$ denote the Bloch eigenstate and eigenvalue, respectively.
Here $\partial_{k_y}H(\bm{k})$ is evaluated in lattice units so that $\chi_{zy}$ is dimensionless; the physical prefactor enters through Eq.~\eqref{eq:edelstein_scaling}.
In the collinear limit ($\theta=0$),  $\mathcal{PT}$-symmetry enforces $\chi_{zy}=0$ . Conversely, the orthogonal $p$-wave phase breaks this symmetry, allowing a finite current-induced spin polarization.

\begin{figure}[t]
\centering
\includegraphics[width=1\linewidth]{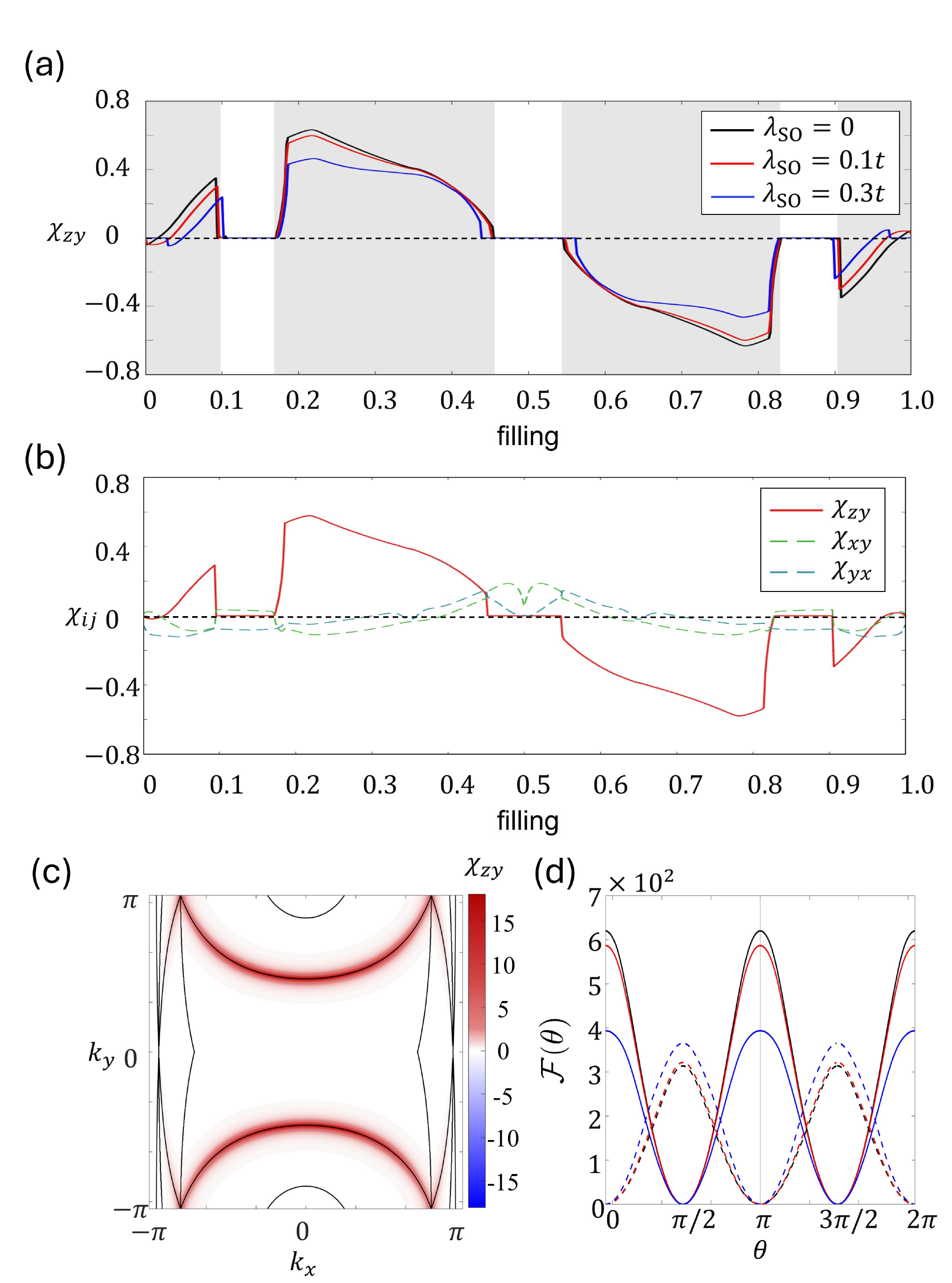}
\caption{Computed from Eq.~(\ref{eq:H_mid_compact}) with the Rashba term and Eq.~(\ref{eq:chi_kernel_def}) with $t=1, m_1=m_2=1, \lambda_{SO}/t=0,0.1,0.3$ as indicated; $\theta$ set to the equilibrium configuration at each filling.(a) Edelstein kernel $\chi_{zy}$ as a function of electronic filling for different SOC strengths. The shaded regions indicate the $p$-wave phase ($\theta \approx \pi/2$) where the response is significant. (b) Filling dependence of representative tensor components $\chi_{ij}$ in the presence of Rashba SOC ($\lambda_{\rm SO}/t=0.1$), showing that transverse components (e.g., $\chi_{xy}$ and $\chi_{yx}$) remain subdominant to $\chi_{zy}$ in the $p$-wave regime. (c) Momentum-resolved kernel of the response, showing the odd-parity distribution over the Fermi surface (black lines). (d) Relative total energy $\mathcal{F}(\theta)$ for representative fillings of $1/2$ (collinear stable) and $1/3$ (orthogonal stable).}
\label{fig:SOC_Edelstein}
\end{figure}

Fig.~\ref{fig:SOC_Edelstein}(a) demonstrates the evolution of $\chi_{zy}$ as a function of electron filling. The response exhibits a sharp contrast driven by the magnetic phase transition: it vanishes in the collinear regimes but becomes finite within the metallic $p$-wave windows ($\theta \simeq \pi/2$). A side-by-side comparison of the band dispersion and $\chi_{zy}$ at representative fillings is given in Sec.~VI of the Supplemental Material, illustrating directly that the response is controlled by the presence or absence of exchange-driven spin splitting. This behavior originates from the $p$-wave spin splitting at the Fermi surface, as visualized by the momentum-resolved kernel in Fig.~\ref{fig:SOC_Edelstein}(c). This gate-tunability is intrinsic to the magnetic texture and robust against relativistic effects, as confirmed by including a Rashba spin-orbit term
$H_{\rm soc}=\lambda_{\rm SO}(\sin k_x\sigma_y - \sin k_y\sigma_x)$.

Fig.~\ref{fig:SOC_Edelstein}(a) shows that strong SOC ($\lambda_{\rm SO}/t=0.3$), comparable to SOC scales expected in layered Fe-based metals~\cite{Borisenko2016SOC, Day2018SOC}, does not alter the qualitative dependence on filling. Rashba SOC can also generate additional transverse components of the response; indeed, for $\lambda_{\rm SO}/t=0.1$ we find finite $\chi_{xy}$ and $\chi_{yx}$ [Fig.~\ref{fig:SOC_Edelstein}(b)], but their magnitudes remain subdominant compared to the magnetically generated $\chi_{zy}$ in the $p$-wave regime. This robustness confirms that the observed Edelstein effect is driven by magnetic symmetry breaking, distinguishing it from the conventional mechanism that relies on SOC.

\section{Conclusion} We have established a microscopic, symmetry-driven route to $p$-wave magnetism in a sAFM/metal/sAFM heterostructure. As a concrete vdW-compatible platform, we propose $\mathrm{GdTe}_3$ as a candidate material, where spin-polarized STM identified a low-temperature commensurate sAFM phase below $\sim 7$~K~\cite{GdTe3AFM,GdTe3monolayer,GdTe3VdW}. Using $a\simeq 3.1$~\AA\ as a representative in-plane lattice constant (e.g., the lattice constant of the metal layer that is commensurate with $\mathrm{GdTe}_3$~\cite{GdTe3VdW}), together with the momentum relaxation time $\tau_{\rm sc}\sim 0.1$--$0.3$~ps~\cite{GdTe3VdW} and an electric field $E_y=0.1$--$1$~kV\,cm$^{-1}$, we estimate $\delta S_z \sim 10^{-3}$–$10^{-2}(\hbar/2)$ per magnetic unit cell. For a magnetic unit-cell area $A_{\rm cell}\sim 4a^2$, this corresponds to a sheet spin density $\delta S_z/A_{\rm cell}\sim (3\times 10^{-3}$–$3\times 10^{-2})(\hbar/2)\mathrm{nm}^{-2}$, which exceeds the recently reported non-relativistic Edelstein scale of the $p$-wave magnet CeNiAsO~\cite{chakraborty2025} by roughly an order of magnitude, and the maximal relativistic Rashba/topological-insulator interface response~\cite{rojassanchez2016} and the non-collinear antiferromagnet LuFeO$_3$~\cite{gonzalez2024non} by approximately two orders of magnitude (see Supplemental Material, Table~II). Also the estimated response remains of order unity over a broad range of the induced Zeeman potential, as detailed in Sec. VI of the Supplemental Material. Since $\delta S_z \propto \tau_{\rm sc}$, optimizing the spacer quality and interfaces can further enhance the signal, making the effect readily measurable in clean heterostructures.

Besides vdW materials, the same design principle applies to epitaxial thin films of layered sAFM, including the iron pnictide family (AFe$_2$As$_2$ with A = Ca, Sr, Ba; RFeAsO with R = rare earth; and NaFeAs), which host robust stripe order on nearly square Fe lattices\cite{FesAFM1,FesAFM2,FesAFM3,FesAFM4,FesAFM5,FesAFM6,FesAFM7,FesAFMReview}. The proposed mechanism is therefore platform across the family of in-plane stripe-AFM materials with $\mathrm{GdTe}_3$ serving as one representative vdW example rather than a unique requirement.
Among available platforms, vdW assembly is particularly appealing: exfoliation and dry-transfer stacking enable deterministic control of twist and registry.

\section{Acknowledgments}
We thank Changmin Lee for fruitful discussions. This work was supported by the National Research Foundation of Korea (NRF) grant funded by the Korea government (MSIT) (Grants No. RS-2025-16070482, RS-2025-25446099, RS2025-03392969, RS-2025-25464760, RS-2025-25431194, RS-2025-25433233, RS-2026-25607155, RS-2026-25519864 for H.U.). This work was also supported by BK21 FOUR (Fostering Outstanding Universities for Research) program through the National Research Foundation (NRF) funded by the Ministry of Education of Korea. G.B.S. acknowledges support from the Australian Research Council (ARC) through Grant No. DP240100168 and the NRF through Grant No. RS-2024-00453943.

\bibliography{PRL_revision_main.bib}
\end{document}